\documentclass[twocolumn,11pt]{article}
\usepackage{times}
\newdimen\proofrulebreadth \proofrulebreadth=.05em
\newdimen\proofdotseparation \proofdotseparation=1.25ex
\newdimen\proofrulebaseline \proofrulebaseline=2ex
\newcount\proofdotnumber \proofdotnumber=3
\let\then\relax
\def\hfi{\hskip0pt plus.0001fil}
\mathchardef\squigto="3A3B
\newif\ifinsideprooftree\insideprooftreefalse
\newif\ifonleftofproofrule\onleftofproofrulefalse
\newif\ifproofdots\proofdotsfalse
\newif\ifdoubleproof\doubleprooffalse
\let\wereinproofbit\relax
\newdimen\shortenproofleft
\newdimen\shortenproofright
\newdimen\proofbelowshift
\newbox\proofabove
\newbox\proofbelow
\newbox\proofrulename
\def\shiftproofbelow{\let\next\relax\afterassignment\setshiftproofbelow\dimen0 }
\def\shiftproofbelowneg{\def\next{\multiply\dimen0 by-1 }%
\afterassignment\setshiftproofbelow\dimen0 }
\def\setshiftproofbelow{\next\proofbelowshift=\dimen0 }
\def\setproofrulebreadth{\proofrulebreadth}

\def\prooftree{
\ifnum  \lastpenalty=1
\then   \unpenalty
\else   \onleftofproofrulefalse
\fi
\ifonleftofproofrule
\else   \ifinsideprooftree
        \then   \hskip.5em plus1fil
        \fi
\fi
\bgroup
\setbox\proofbelow=\hbox{}\setbox\proofrulename=\hbox{}%
\let\justifies\proofover\let\leadsto\proofoverdots\let\Justifies\proofoverdbl
\let\using\proofusing\let\[\prooftree
\ifinsideprooftree\let\]\endprooftree\fi
\proofdotsfalse\doubleprooffalse
\let\thickness\setproofrulebreadth
\let\shiftright\shiftproofbelow \let\shift\shiftproofbelow
\let\shiftleft\shiftproofbelowneg
\let\ifwasinsideprooftree\ifinsideprooftree
\insideprooftreetrue
\setbox\proofabove=\hbox\bgroup$\displaystyle 
\let\wereinproofbit\prooftree
\shortenproofleft=0pt \shortenproofright=0pt \proofbelowshift=0pt
\onleftofproofruletrue\penalty1
}

\def\eproofbit{
\ifx    \wereinproofbit\prooftree
\then   \ifcase \lastpenalty
        \then   \shortenproofright=0pt  
        \or     \unpenalty\hfil         
        \or     \unpenalty\unskip       
        \else   \shortenproofright=0pt  
        \fi
\fi
\global\dimen0=\shortenproofleft
\global\dimen1=\shortenproofright
\global\dimen2=\proofrulebreadth
\global\dimen3=\proofbelowshift
\global\dimen4=\proofdotseparation
\global\count255=\proofdotnumber
$\egroup  
\shortenproofleft=\dimen0
\shortenproofright=\dimen1
\proofrulebreadth=\dimen2
\proofbelowshift=\dimen3
\proofdotseparation=\dimen4
\proofdotnumber=\count255
}

\def\proofover{
\eproofbit 
\setbox\proofbelow=\hbox\bgroup 
\let\wereinproofbit\proofover
$\displaystyle
}%
\def\proofoverdbl{
\eproofbit 
\doubleprooftrue
\setbox\proofbelow=\hbox\bgroup 
\let\wereinproofbit\proofoverdbl
$\displaystyle
}%
\def\proofoverdots{
\eproofbit 
\proofdotstrue
\setbox\proofbelow=\hbox\bgroup 
\let\wereinproofbit\proofoverdots
$\displaystyle
}%
\def\proofusing{
\eproofbit 
\setbox\proofrulename=\hbox\bgroup 
\let\wereinproofbit\proofusing
\kern0.3em$
}

\def\endprooftree{
\eproofbit 
  \dimen5 =0pt
\dimen0=\wd\proofabove \advance\dimen0-\shortenproofleft
\advance\dimen0-\shortenproofright
\dimen1=.5\dimen0 \advance\dimen1-.5\wd\proofbelow
\dimen4=\dimen1
\advance\dimen1\proofbelowshift \advance\dimen4-\proofbelowshift
\ifdim  \dimen1<0pt
\then   \advance\shortenproofleft\dimen1
        \advance\dimen0-\dimen1
        \dimen1=0pt
        \ifdim  \shortenproofleft<0pt
        \then   \setbox\proofabove=\hbox{%
                        \kern-\shortenproofleft\unhbox\proofabove}%
                \shortenproofleft=0pt
        \fi
\fi
\ifdim  \dimen4<0pt
\then   \advance\shortenproofright\dimen4
        \advance\dimen0-\dimen4
        \dimen4=0pt
\fi
\ifdim  \shortenproofright<\wd\proofrulename
\then   \shortenproofright=\wd\proofrulename
\fi
\dimen2=\shortenproofleft \advance\dimen2 by\dimen1
\dimen3=\shortenproofright\advance\dimen3 by\dimen4
\ifproofdots
\then
        \dimen6=\shortenproofleft \advance\dimen6 .5\dimen0
        \setbox1=\vbox to\proofdotseparation{\vss\hbox{$\cdot$}\vss}%
        \setbox0=\hbox{%
                \advance\dimen6-.5\wd1
                \kern\dimen6
                $\vcenter to\proofdotnumber\proofdotseparation
                        {\leaders\box1\vfill}$%
                \unhbox\proofrulename}%
\else   \dimen6=\fontdimen22\the\textfont2 
        \dimen7=\dimen6
        \advance\dimen6by.5\proofrulebreadth
        \advance\dimen7by-.5\proofrulebreadth
        \setbox0=\hbox{%
                \kern\shortenproofleft
                \ifdoubleproof
                \then   \hbox to\dimen0{%
                        $\mathsurround0pt\mathord=\mkern-6mu%
                        \cleaders\hbox{$\mkern-2mu=\mkern-2mu$}\hfill
                        \mkern-6mu\mathord=$}%
                \else   \vrule height\dimen6 depth-\dimen7 width\dimen0
                \fi
                \unhbox\proofrulename}%
        \ht0=\dimen6 \dp0=-\dimen7
\fi
\let\doll\relax
\ifwasinsideprooftree
\then   \let\VBOX\vbox
\else   \ifmmode\else$\let\doll=$\fi
        \let\VBOX\vcenter
\fi
\VBOX   {\baselineskip\proofrulebaseline \lineskip.2ex
        \expandafter\lineskiplimit\ifproofdots0ex\else-0.6ex\fi
        \hbox   spread\dimen5   {\hfi\unhbox\proofabove\hfi}%
        \hbox{\box0}%
        \hbox   {\kern\dimen2 \box\proofbelow}}\doll%
\global\dimen2=\dimen2
\global\dimen3=\dimen3
\egroup 
\ifonleftofproofrule
\then   \shortenproofleft=\dimen2
\fi
\shortenproofright=\dimen3
\onleftofproofrulefalse
\ifinsideprooftree
\then   \hskip.5em plus 1fil \penalty2
\fi
}

\newtheorem{definition}{Definition}

\begin{document}
\global\def\refname{{\normalsize \it References:}}
\baselineskip 12.5pt
%
%
%
\title{\LARGE \bf ImpNet: Programming Software-Defied Networks Using
Imperative Techniques\thanks{This is an extended and revised version
of~\cite{El-Zawawy14}.}}

\date{}

\author{\hspace*{-10pt}
\begin{minipage}[t]{3.4in} \normalsize \baselineskip 12.5pt
\centerline{Mohamed A. El-Zawawy$^{1,2,}$\footnote{Corresponding
author.}}\centerline{$^1$College of Computer and Information
Sciences} \centerline{Al Imam Mohammad Ibn Saud Islamic University}
\centerline{ (IMSIU)} \centerline{Riyadh }\centerline{ Kingdom of
Saudi Arabia}\centerline{} \centerline{$^2$Department of
Mathematics} \centerline{Faculty of Science} \centerline{Cairo
University} \centerline{Giza 12613}
\centerline{Egypt}\centerline{maelzawawy@cu.edu.eg}
\end{minipage} \kern 0in
\begin{minipage}[t]{3.3in} \normalsize \baselineskip 12.5pt
\centerline{Adel I. AlSalem} \centerline{College of Computer and
Information Sciences} \centerline{Al Imam Mohammad Ibn Saud Islamic
University} \centerline{ (IMSIU)}  \centerline{Riyadh }\centerline{
Kingdom of Saudi Arabia} \centerline{alsalem@ccis.imamu.edu.sa}
\end{minipage}
%
%
\\ \\ \hspace*{-10pt}
\begin{minipage}[b]{6.9in} \normalsize
\baselineskip 12.5pt {\it Abstract:} Software and hardware
components are basic parts of modern networks. However the software
component is typical sealed and function-oriented. Therefore it is
very difficult to modify these components. This badly affected
networking innovations. Moreover, this resulted in network policies
having complex interfaces that are not user-friendly and hence
resulted in huge and complicated flow tables on physical switches of networks.
This greatly degrades the network performance in many cases.   \\
Software-Defined Networks (SDNs) is a modern architecture of
networks to overcome issues mentioned above. The idea of SDN is to
add to the network a controller device that manages all the other
devices on the network including physical switches of the network.
One of the main tasks of the managing process is \textit{switch
learning}; achieved via programming physical switches of the network
by adding or removing rules for packet-processing to/from switches,
more
specifically to/from their flow tables.\\
A high-level imperative network programming language, called
\textit{ImpNet}, is presented in this paper. \textit{ImpNet} enables
writing efficient, yet simple, and powerful programs to run on the
controller to control all other network devices including switches.
\textit{ImpNet} is compositional, simply-structured, expressive, and
more importantly imperative. The syntax of \textit{ImpNet} together
two types of operational semantics to contracts of \textit{ImpNet}
are presented in the paper. The proposed semantics are of the static
and dynamic types. Two modern application programmed using
\textit{ImpNet} are shown in the paper as well. The semantics of the
applications are shown in the paper also.
\\ [4mm] {\it Key--Words:}
Network programming languages, Controller-switch architecture,
Static operational semantics, Dynamic operational semantics, Syntax,
\textit{ImpNet}, Software-defined networks.
\end{minipage}
\vspace{-10pt}}

\maketitle

\thispagestyle{empty} \pagestyle{empty}
%
%

\section{Introduction}
\label{intro}\vspace{-4pt}

A network is a group of appliances connected to exchange data. Among
these appliances are switches forwarding data depending on MAC
addresses, routers forwarding data depending on IP addresses, and
firewalls taking care of forbidden data. The network appliances are
connected using a model that efficiently allows  forwarding,
storing, ignoring, tagging, and providing statistics about data
moving in the network. Some of the network appliances, like
routers~\cite{Arneson,Suzuki}, are special in their functionality as
they have some control over the network. This enables routers to
compute and determine routes of data in the network. Of course
different networks have different characteristics and abilities.

In 2011, the Open Networking Foundation~\cite{website}, suggested
removing the control owned by different network appliances and
adding, instead, a general-purpose appliance, controller, to program
different network appliances and querying data flowing in the
network. The impact of this simple suggestion is huge; giant
networks do not need special-purpose, complex, expensive switches
any more. In such networks, cheap programmable switches can be used
and programmed to configure and optimize networks via writing
programs~\cite{Rexford} running on controllers.

Software-Defined Networks (SDNs)~\cite{Foster13} are networks
established using the controller-switch architecture. A precise
implementation of this architecture is OpenFlow~\cite{Cai10} used to
achieve various network-wide applications such as monitoring data
flow, balancing switch load, network management, controlling
appliances access, detection of service absence, host mobility, and
forwarding data center. Therefore SDNs caused the appearance of
network programming languages~\cite{Monsanto,Hong,Elsts,Bain}.

This paper presents \textit{ImpNet}, an imperative high-level
network programming language. \textit{ImpNet}  expresses commands
enabling controllers to program other network appliances including
switches. \textit{ImpNet} has a clear and simply-structured syntax
based on classical concepts of imperative programming that allow
building rich and robust network applications in a natural way.
\textit{ImpNet} can be realized as a generalization of
Frenetic~\cite{Foster10} which is a functional network programming
language. This is clear by the fact that the core of programs
written in \textit{ImpNet} and Frenetic is based on a query result
in the form of stream of values (packets, switches IDs, etc.).
Commands for treating packets in \textit{ImpNet} include
constructing and installing (adding to flow tables of switches)
switch rules. \textit{ImpNet} supports building simple programs to
express complex dynamic functionalities like load balancing and
authentication. \textit{ImpNet} programs can also analyze packets
and historical traffic patterns.

The current paper presents both syntax and semantics of
\textit{ImpNet}. Actually two types of precise operational semantics
for \textit{ImpNet} are presented in this paper; static and
semantics. Dynamic semantics\footnote{Dynamic semantics can be
realized as a perspective on programming languages semantics that
models the increase of data in time.} are very useful for studying
and analyzing programs at run time. Such semantics have a wide range
of applications in case of network programs.  This is so as the
network programs tend to be event-driven and hence their run times
are much longer than that of many other application-programs.
Therefore a formal definitions for run times behaviors (semantics)
can easily be employed to achieve runtime verifications for
controller programs. Moreover two \textit{ImpNet} programs achieving
two important controller applications are also presented in this
paper together with their precise operational semantics.

The current paper is an extended and revised version
of~\cite{El-Zawawy14}. The current paper extends the work
of~\cite{El-Zawawy14} by supporting the theoretical foundations
\textit{ImpNet} via a dynamic operation semantics for its programs.
The dynamic semantics is necessary for achieving many important
dynamics verifications for network programs. The extensions are
mainly included in Section~\ref{s1n} and Figures~\ref{dyn1}
and~\ref{dyn2}.

\begin{figure*}[t]
\centering \fbox{
\begin{minipage}{12 cm}
{\footnotesize{
\begin{eqnarray*}
& &x\in \hbox{lVar}\qquad\qquad Q\in \hbox{Queries}\qquad\qquad n\in \hbox{Integers}\\
{et} \in \hbox{Eventrans}   &::= & n\mid {\hbox{Lift} (x,\lambda
t.f(t))}\mid {\hbox{ApplyLft} (x,\lambda t.f(t))}\mid
{\hbox{ApplyRit}
(x,\lambda t.f(t))}\mid \\
&& \hbox{Merge}(x_1,x_2)\mid
\hbox{MixFst}(A,x_2,x_3)\mid\hbox{MixSnd}(A,x_2,x_3)\mid \\ & &
\hbox{Filter}(x,\lambda.f(t))\mid \hbox{Once}(x)
\mid \hbox{MakForwRule}(x)\mid\hbox{MakeRule}(x)\\
S \in \hbox{Stmts}   &::= &  {x:= {et}}\mid S_1; S_2\mid
\hbox{AddRules}(x)\mid \hbox{Register}\mid \hbox{Send}(x)\\ & &
\hbox{If }(x)\ \hbox{then }S_1\hbox{ else }S_2\mid \hbox{While }(x)\
\hbox{do }S
\\
D\in \hbox{Defs}   &::= & \epsilon\mid {x:= Q}\mid D D.
\\
p\in \hbox{Progs}   &::= & D\gg S.
\end{eqnarray*}
}}\caption{ImpNet Syntax.}\label{f1}
\end{minipage}
}
\end{figure*}

\subsection*{Motivation}
\vspace{-4pt} The motivation of this paper is the lack of a simple
syntax for an imperative network programming language. Yet, a
stronger motivation is that most existing network programming
languages are not supported theoretically (using static operational
semantics, dynamic operational semantics, type systems, program
logics like Floyd--Hoare logic, etc.).

\subsection*{Contributions}
\vspace{-4pt}

Contributions of this paper are the following.
\begin{enumerate}
\item  A new simply-structured syntax for an imperative network
programming language; \textit{ImpNet}.
\item A static operational semantics (in the form of states and inference rules)
for constructs of \textit{ImpNet}.
\item A dynamic operational semantics for constructs of \textit{ImpNet}.
\item Two detailed examples of programs constructed in \textit{ImpNet}
with their precise operation semantics.
\end{enumerate}

\subsubsection*{Organization}
\vspace{-4pt}

The rest of this paper is organized as following. Section~\ref{s3}
reviews related work. Section~\ref{s1} presents the syntax and
static operational semantics of \textit{ImpNet}. The proposed static
semantics are operational and hence consists of states and inference
rules presented in Section~\ref{s1}. A dynamic operational semantics
of \textit{ImpNet} programs is introduced in Section~\ref{s1n}. Two
detailed examples of programmes built in \textit{ImpNet} are
presented in Section~\ref{s2}. This section also explains how the
two examples can be assigned precise static operational semantics
using our proposed static operational semantics. Section~\ref{s3a}
gives directions for future work and Section~\ref{s4} concludes the
paper.

\section{Related Work}
\label{s3}\vspace{-4pt}

This section presents work most related to that presented in the
current paper. One of the early attempts to develop software-defined
networking (SDN) is NOX~\cite{Gude08} based on ideas
from~\cite{Casado09} and 4D~\cite{Greenberg05}. On the switch-level,
NOX uses explicit and callbacks rules for packet-processing.
Examples of applications that benefitted from NOX are load
balancer~\cite{Wang11} and the work in~\cite{Handigol09,Heller10}.
Many directions for improving platforms of programming networks
include Maestro~\cite{Cai10} and Onix~\cite{Koponen10}, which uses
distribution and parallelization to provide better performance and
scalability.

A famous programming language for networks is
Frenetic~\cite{Foster10,Foster11} which has two main components. The
first component is a collection of operators that are source-level.
The operators aim at establishing and treating streams of network
traffic. These operators also are built on concepts of functional
programming (FP) and query languages of declarative database.
Moreover the operators support a modular design, a cost control, a
race-free semantics, a single-tier programming, and a declarative
design. The second component of Frenetic is a run-time system. This
system facilitates all of the actions of adding and removing
low-level rules to and from flow tables of switches. One advantage
of \textit{ImpNet}, the language presented in this paper, over
Frenetic is that \textit{ImpNet} is imperative. Therefore
\textit{ImpNet} paves the way to the appearance of other types of
network programming languages such as object-oriented network
programming langues and context-oriented network programming
languages.

Other examples to program network components though high-level
languages are NDLog and NetCore~\cite{Loo05netcore}. NetCore
provides an integrated view of the whole network. NDLog is designed
in an explicitly distributed fashion.

As an extension of Datalog, NDLog~\cite{Loo05,Loo05-2} was presented
to determine and code protocols of routing~\cite{Arneson}, overlay
networks, and concepts like hash tables of  distributed systems.
\textit{ImpNet} (presented in this paper), Frenetic, and NDLog can
be classified as high-level network programming languages. While
NDLog main focus is overlay networks and routing protocols, Frenetic
(in a functional way) and \textit{ImpNet} (in an imperative way)
focus on implementing packet processing such as modifying header
fields. Therefore \textit{ImpNet} equips a network programmer with a
modular view of the network which is not provided by NDLog and
Frenetic. This is supported by the fact that  a program in NDLog is
a single query that is calculated on each router of the network. One
advatnage of network programming langauges (\textit{ImpNet}) is
saving routing energy~\cite{Bhanumathi12}.

Energy Efficient Routing with Transmission Power Control based
Biobjective Path Selection Model for Mobile Ad-hoc Network

The switch component~\cite{McKeown} of networks can be programmed
via many interfaces such as OpenFlow platform. Examples of other
platforms include Shangri-La~\cite{Chen} and FPL-3E~\cite{Cristea},
RouteBricks~\cite{Dobrescu}, Click modular router~\cite{Kohler},
Snortran~\cite{Egorov} and Bro~\cite{Paxson}. The idea in
Shangri-La~\cite{Chen} and FPL-3E~\cite{Cristea} is to produce
certain hardware for packet-processing from high-level programs that
achieves packet-processing. In RouteBricks~\cite{Dobrescu}, stock
machines are used to improve performance of program switches. As a
modular approach, the platform of Click modular
router~\cite{Kohler}, enables programming network components.  This
system focuses on software switches in the form of Linux kernel
code. For the sake of intrusions detection and preserving network
security, Snortran~\cite{Egorov} and Bro~\cite{Paxson} enable coding
monitoring strategies and robust packet-filtering. One advantage of
\textit{ImpNet}, the language presented in this paper, over all the
related work is that \textit{ImpNet} overcomes the disadvantage of
most similar languages of focusing on controlling a single device.

There are many possible network applications for dynamics semantics
of \textit{ImpNet}.  The K-random search in peer-to-peer networks
using dynamic semantic data replication~\cite{Cao12} can be better
verified using dynamic semantics of the nature proposed in this
paper. The correctness of dynamic information retrieval in P2P
networks is achieved using dynamics semantics
platform~\cite{Eftychiou12}. Dynamic reconfiguration of networked
services can be carried out using dynamic semantic in the shape of
interoperability framework~\cite{Kamoun12} like that of this paper.
In an attractor network, spreading activation with latching dynamics
can be realized using automatic dynamic semantic similar to the one
proposed in this paper~\cite{Lerner12}.

\section{Syntax and Static Operational Semantics}\label{s1}\vspace{-4pt}

This section presents the syntax and static operational semantics of
\textit{ImpNet}, a high-level programming language for SDN networks
using the switch-controller architecture. Figure~\ref{f1} shows the
syntax of \textit{ImpNet}. Figures~\ref{f2} and~\ref{f3} present the
static operational semantics of \textit{ImpNet} constructs. The
proposed semantics is operational and its states are defined in the
following definition.

\begin{definition}\label{state}
\begin{enumerate}
\item $t\in$ Types = \{int, Switch IDs, Packet,
(Switch IDs, int, bool)\}$\cup\{(t_1,t_2)\mid t_1,t_2\in
\hbox{Types}\}$.
\item $v\in\hbox{Values} =$ Natural numbers $\ \cup \hbox{ Switch IDs }
 \cup\hbox{Packets}\cup \hbox{Switch IDs }\times $ Natural numbers $\times $
 Boolean values\ $\cup\
 \{(v_1,v_2)\mid v_1,v_2\in \hbox{Values}\}$.
The expression $v:t$ denotes that the type of the value $v$ is $t$.
\item ${ev}\in \hbox{Events} = \{(v_1,v_2,\ldots,v_n)\mid
\exists t (\forall i\ v_i:t)\}$.
\item Actions= {\{sendcontroller, sendall, sendout, change(h,v) \}}.
\item $r\in\hbox{Rules} =\hbox{Patterns} \times \hbox{Acts}$.
\item ${rl}\in \hbox{RlLst}=\{[r_1,r_2,\ldots,r_n]\mid r_i\in\hbox{Rules}\}$.
\item ${ir}\in \hbox{Intial-rule-assignment}=\hbox{SwchIds}\times \hbox{Rules}$.
\item $\sigma\in\hbox{SwchSts}=\hbox{Flow-tables} =\hbox{SwchIds}
\rightarrow \hbox{RlLst}$.
\item $\gamma\in\hbox{VarSts} =\hbox{Var} \rightarrow
\hbox{Events}\cup \hbox{RlLst} $.
\item $s\in \hbox{States} =\hbox{SwchSts}\times
\hbox{VarSts}\times\hbox{RlLst}. $
\end{enumerate}
\end{definition}

A program in \textit{ImpNet} is a sequence of queries followed by a
statement. The result of each query is an event which is a finite
sequence of values. The event concept is also used in Frenetic.
However an event in Frenetic is an infinite sequence of values. A
value is an integer, a switch ID, a packet, a triple of a switch ID,
an integer, and a Boolean value, or a pair of two values. Each value
has a type of the set \textit{Types}. In this paper, we focus on the
details of statements as this is the most interesting part in a
network programming language.

The query part of a network language is there to enable reading the
status of the network. Typically, queries contain commands for
\begin{itemize}
    \item dividing packets via grouping according to values of header fields,
    \item splitting packets according to arrival time or values of header fields,
    \item filtering packets in the network according to a given pattern,
    \item minimizing the volume of returned values, and
    \item summarizing results using size or number of packets.
\end{itemize}

Possible actions taken by a certain switch on a certain packet are
\textit{sendcontroller}, \textit{sendall}, \textit{sendout}, or
\textit{change(h,v)}. The action \textit{sendcontroller} sends a
packet to the controller to process it. The action \textit{sendall}
sends the packet to all other switches. The action \textit{sendout}
sends the packet out of the switch through a certain port. The
action \textit{change(h,v)} modifies the header field $h$ of the
packet to the new value $v$.

\begin{figure*}
\centering \fbox{
\begin{minipage}{15cm}
{\footnotesize{
\[
\begin{prooftree}
v_i:t \qquad \gamma(x)=(v_1,v_2,\ldots,v_n)\justifies \hbox{Lift}
(x,\lambda t.f(t)):\gamma\rightarrow
(f(v_1),f(v_2),\ldots,f(v_n))\thickness=0.08em\using{(\hbox{Lift}^s)}
\end{prooftree}
\]
\[
\begin{prooftree}
\gamma(x_1)=(v_1,v_2,\ldots,v_n)\qquad
\gamma(x_2)=(w_1,w_2,\ldots,w_n)\justifies
\hbox{Merge}(x_1,x_2):\gamma\rightarrow
((v_1,w_1),(v_2,w_2),\ldots,(v_n,w_n))\thickness=0.08em\using{(\hbox{Merge}^s)}
\end{prooftree}
\]
\[
\begin{prooftree}
\gamma(x)=(v_1,v_2,\ldots,v_n)\qquad A=\{i\mid f(v_i)=\hbox{true}\}
\justifies \hbox{Filter}(x,\lambda.f(t)):\gamma\rightarrow
(\dots,v_i,\ldots\mid i\in A)
\thickness=0.08em\using{(\hbox{Filter}^s)}
\end{prooftree}
\]
\[
\begin{prooftree}
v_i:t\qquad
\gamma(x)=((v_1,v_1^\prime),(v_2,v_2^\prime),\ldots,(v_n,v_n^\prime))\justifies
\hbox{ApplyLft} (x,\lambda t.f(t)):\gamma\rightarrow
((f(v_1),v_1^\prime),(f(v_2),v_2^\prime),\ldots,(f(v_n),v_n^\prime))\thickness=0.08em
\using{(\hbox{App}_1^s)}
\end{prooftree}
\]
\[
\begin{prooftree}
v_i^\prime:t\qquad
\gamma(x)=((v_1,v_1^\prime),(v_2,v_2^\prime),\ldots,(v_n,v_n^\prime))\justifies
\hbox{ApplyRit} (x,\lambda t.f(t)):\gamma\rightarrow
((v_1,f(v_1^\prime)),(v_2,f(v_2^\prime)),\ldots,(v_n,f(v_n^\prime)))\thickness=0.08em\using{(\hbox{App}_2^s)}
\end{prooftree}
\]
\[
\begin{prooftree}
\hbox{type}(x)\in \hbox{Types}
\justifies\hbox{Once}(x):\gamma\rightarrow
(\underbrace{x,x,\ldots,x}_{n\_times})
\thickness=0.08em\using{(\hbox{Once}^s)}
\end{prooftree}
\]
\[
\begin{prooftree}
\gamma(x_1)=(v^1_1,v^1_2,\ldots,v^1_n)\quad
\gamma(x_2)=(v^2_1,v^2_2,\ldots,v^2_n)\quad
A_1=A\cup\{v^1_1\}\quad\forall i>1. A_i=A_{i-1}\cup \{v^1_i\}
\justifies \hbox{MixFst}(A,x_1,x_2):\gamma\rightarrow
((A_1,v^2_1),(A_2,v^2_2),\ldots,(A_n,v^2_n))
\thickness=0.08em\using{(\hbox{Mix}_1^s)}
\end{prooftree}
\]
\[
\begin{prooftree}
\gamma(x_1)=(v^1_1,v^1_2,\ldots,v^1_n)\quad
\gamma(x_2)=(v^2_1,v^2_2,\ldots,v^2_n)\quad
A_1=A\cup\{v^2_1\}\quad\forall i>1. A_i=A_{i-1}\cup \{v^2_i\}
\justifies \hbox{MixSnd}(A,x_1,x_2):\gamma\rightarrow
((v^1_1,A_1),(v^1_2,A_2),\ldots,(v^1_n,A_n))
\thickness=0.08em\using{(\hbox{Mix}_2^s)}
\end{prooftree}
\]
\[
\begin{prooftree}
\gamma(x)=((v^1_1,v^2_1,v^3_1),(v^1_2,v^2_2,v^3_2),\ldots,(v^1_n,v^2_n,v^3_n))
\justifies \hbox{MakForwRule}(x):\gamma\rightarrow
[(v^1_1,(v^3_1,sendout(v^2_1))),(v^1_2,(v^3_2,sendout(v^2_2))),
\dots,(v^1_n,(v^3_n,sendout(v^2_n)))]
\thickness=0.08em\using{(\hbox{MFR}^s)}
\end{prooftree}
\]
\[
\begin{prooftree}
\gamma(x)=((v^1_1,a_1,v^2_1),(v^1_2,a_2,v^2_2),\ldots,(v^1_n,a_n,v^2_n))
\justifies \hbox{MakeRule}(x):\gamma\rightarrow
[(v^1_1,a_1(v^1_2)),(v^2_1,a_2(v^2_2)),\dots,(v^i_n,a_n(v^n_2))]
\thickness=0.08em\using{(\hbox{MkRl}^s)}
\end{prooftree}
\]
}} \caption{Operational Semantics for Event Functions of
ImpNet}\label{f2}
\end{minipage}}\\ \fbox{
\begin{minipage}{15cm}
{\footnotesize{
\[
\begin{prooftree}
et:\gamma\rightarrow u \justifies x:=
{et}:(\sigma,\gamma,ir)\rightarrow(\sigma,\gamma[x\mapsto u],ir)
\thickness=0.08em\using{(\hbox{Assgn}^s)}
\end{prooftree}
\]
\[
\begin{prooftree}
S_1:(\sigma,\gamma,ir)\rightarrow
(\sigma^{\prime\prime},\gamma^{\prime\prime},ir^{\prime\prime})\qquad
S_2: (\sigma^{\prime\prime},\gamma^{\prime\prime},ir^{\prime\prime})
\rightarrow(\sigma^\prime,\gamma^\prime,ir^\prime) \justifies S_1;
S_2:(\sigma,\gamma,ir)\rightarrow(\sigma^\prime,\gamma^\prime,ir^\prime)
\thickness=0.08em\using{(\hbox{seq}^s)}
\end{prooftree}
\]
\[
\begin{prooftree}
\gamma(x)\in \hbox{Intial-rule-assignment}\justifies
\hbox{AddRules}(x):(\sigma,\gamma,ir)\rightarrow(\sigma,\gamma,ir\cup
\gamma(x) ) \thickness=0.08em\using{(\hbox{Addrl}^s)}
\end{prooftree}
\]
\[
\begin{prooftree}
\justifies \hbox{Register}:(\sigma,\gamma,ir)\rightarrow(\sigma\cup
ir,\gamma,\emptyset) \thickness=0.08em\using{(\hbox{Reg}^s)}
\end{prooftree}
\]
\[
\begin{prooftree}
\gamma(x)=((v^1_1,v^2_1,v^3_1),(v^1_2,v^2_2,v^3_2),\ldots,(v^1_n,v^2_n,v^3_n))
\qquad \forall i. (v^i_2,v^i_3)\in \hbox{history}(v^i_1) \justifies
\hbox{Send}(x):(\sigma,\gamma,ir)\rightarrow(\sigma,\gamma,ir)
\thickness=0.08em\using{(\hbox{Send}^s)}
\end{prooftree}
\]}}
\caption{Operational Semantics for Statements of ImpNet}\label{f3}
\end{minipage}
}\caption{Static Operational Semantics for ImpNet.}\label{f23}
\end{figure*}

A rule in our static operational semantics is a pair of
\textit{pattern} and \textit{action} where \textit{pattern} is a
form that concretely describes a set of packets and \textit{action}
is the action to be taken on elements of this set of packets. Rules
are stored in tables (called \textit{flow tables}) of switches.
\textit{Intial-rule-assignment} represents an initial assignment of
rules to flow tables of switches.

A state in the proposed static operational semantics is a triple
$(\sigma,\gamma,ir)$. In this triple $\gamma$ captures the current
state of the program variables and hence is a map from the set of
variables to the set of events and rule lists. This is so because in
\textit{ImpNet} variables may contain events or rule lists. The
symbol $\sigma$ captures the current state of flow tables of
switches and hence is a map from  \textit{Switche IDs} to rule
lists. Finally, $ir$ is an initial assignment of rules assigned to
switches but have not been registered yet (have not been added to
$\gamma$ yet).

There are five type of statements in \textit{ImpNet}. The assignment
statement $x:= {ef}$ assigns the result of an event transformer (et)
to the variable $x$. The statement \textit{AddRules(x)} adds the
switch rules stored in $x$ to the reservoir of initially assigned
rules. These are rules that are assigned to switches but are not
added to flow tables yet. The statement \textit{Register} makes the
initial assignments permeant by adding them to flow tables of
switches. The statement \textit{Send(x)} sends specific packets to
be treated in a certain way at certain switches. To keep a record of
of actions taken on packets on different switches we assume a map
called \textit{history} from the set of switche IDs to the set of
lists of pairs of packets and taken actions. This map is used in the
Rule $(\hbox{Send}^s)$. Static operational semantics of these
statements are given in Figure~\ref{f3}. Judgments of inference
rules in this figure have the form
$S:(\sigma,\gamma,ir)\rightarrow(\sigma^\prime,\gamma^\prime,ir^\prime)$.
This judgement reads as following. If the execution of $S$ in the
state $(\sigma,\gamma,ir)$ ends then the execution reaches the state
$(\sigma^\prime,\gamma^\prime,ir^\prime)$.

Inference rules in Figure~\ref{f3} use that in Figure~\ref{f2} to
get the semantics of the other important construct of
\textit{ImpNet} which is event transformers (et). Judgements of
Figure~\ref{f2} have the form $et:\gamma\rightarrow u$ meaning that
the semantics of the transformer $et$ in the variable state $\gamma$
is $u$. The event transformer  $\hbox{Lift} (x,\lambda t.f(t))$
applies the map $\lambda t.f(t)$ to values of the event in $x$ (Rule
$(\hbox{Lift}^s)$). The event transformer
$\hbox{Filter}(x,\lambda.f(t))$ filters the event in $x$ using the
map $\lambda t.f(t)$ (Rule $(\hbox{Filter}^s)$). From a given set of
actions $A$ and two events $x_1$ and $x_2$ the event transformers
$\hbox{MixFst}(A,x_1,x_2)$ and $\hbox{MixSnd}(A,x_1,x_2)$ create
lists of rules (Rules $(\hbox{Mix}_1^s)$ and $(\hbox{Mix}_2^s)$).

\section{Dynamic Semantics}\label{s1n}\vspace{-4pt}

This section presents a dynamic semantics for constructs of
\textit{ImpNet} using Rewriting Logic Semantics~\cite{Serbanuta09}.
An important concept to present the semantics is that of rewrite
theories
$(\prod_\textit{ImpNet},E_\textit{ImpNet},R_\textit{ImpNet})$. This
concept uses an extended version of the langauge syntax
$\prod_\textit{ImpNet}$, a group of equations, $E_\textit{ImpNet}$,
built on $\prod_\textit{ImpNet}$, and  set of rules,
$R_\textit{ImpNet}$, for $\prod_\textit{ImpNet}$ constructs. Via a
set of rearrangements, $E_\textit{ImpNet}$ is necessary to prepare
the environment for applying the rules and hence $E_\textit{ImpNet}$
expresses no computational semantics. However $R_\textit{ImpNet}$ is
the semantics component modeling in an irreversible way the
computations. All in all, the semantics uses rewrite theories to
define \textit{ImpNet}. This semantics is built on equational
logic~\cite{Harding13} and hence it is allowed for terms to replace
equal terms in any context. This adds to the power of the semantics.
The idea is that equations are meant to be applied until arriving at
a term matching the l.h.s. of one of the rules. This rule can then
be used to do an irreversible transformation to the term. This type
of semantics is typically efficiently executable.

The proposed semantics uses a modular framework named $C$. In this
framework rules use only necessary configuration items. Hence
configuration changes,  e.g. adding stacks, do not imply modifying
existing rules. Sequences, maps, and bags are necessary concepts to
definitions of the $C$ language. This paper uses the classical
interception of these concepts as data-structures of the equational
type. Sequences, maps, and bags are denoted by
$\hbox{Sq}_e^{\_\hookrightarrow\_},\
\hbox{Mp}_e^{\_\hookrightarrow\_}, \hbox{ and
}\hbox{Bg}_e^{\_\hookrightarrow\_}$, respectively, where
$\hookrightarrow$ is a binary operator and $e$ is the unit element.
Therefore in our proposed dynamic semantics an environment (a state)
is a finite bag of pairs. The domain of a function can easily be
expressed as a bag of items.

Definition~\ref{dynamic-state} gives a formal presentation of states
(configurations) of the proposed dynamic semantics.

\begin{definition}\label{dynamic-state}
\begin{itemize}
    \item $C^s=C\mid \hbox{Sq}_.^{\_\hookrightarrow\_}[C]$.
    \item $\sigma\in\hbox{SwchSts}=\hbox{Mp}_.^{\_,\_}[\hbox{SwchIds},
    \hbox{RlLst}]$.
     \item $h\in\hbox{History}=\hbox{Mp}_.^{\_,\_}[\hbox{SwchIds},
    [\hbox{Packets},\hbox{Acts}]]$.
    \item $\gamma\in\hbox{VarSts}=\hbox{Mp}_.^{\_,\_}[\hbox{Var},
    \hbox{Events}\cup\hbox{RlLst}]$.
    \item $\hbox{confitm}\in\hbox{Configurations items}= <C>_{C^s}\ \mid\
    {<\sigma>_{\hbox{SwchSts}}}
    \ \mid\ {<\gamma>_{\hbox{VarSts}}}\ \mid\  {<rl>_{\hbox{RlLst}}}\
    \mid\  {<h>_{\hbox{History}}}.$
    \item $\hbox{conf}\in\hbox{Configurations}=<\hbox{Bg}_.^{\_}[\hbox{confitm}]>$.
\end{itemize}
\end{definition}

The abstract syntax of \textit{ImpNet}, history maps, maps,
sequences, and bags are used as configuration constructors. The
configuration of \textit{ImpNet} include five
components:\begin{itemize}
    \item  $<C>_{C^s}$ including the computations,
    \item  ${<\sigma>_{\hbox{SwchSts}}}$ capturing flow tables of
the physical switch of the concerned network,
    \item ${<\gamma>_{\hbox{VarSts}}}$ holding the mapping for the program
variables including event variables,
\item ${<rl>_{\hbox{RlLst}}}$
including the list of rules to be registered, and finally
\item ${<h>_{\hbox{History}}}$ capturing the history of packets treatment
of the network.
\end{itemize}

Algebraic structures in definitions of configurations achieve the
context-sensitivity  in $C$ definitions. The sequentialization
operator $\hookrightarrow$ in the dynamic semantics also contributes
to the context-sensitivity.

\begin{figure*}
\centering \fbox{
\begin{minipage}{12 cm}
{\footnotesize{
\begin{eqnarray*}
& &x\in \hbox{lVar}\qquad\qquad Q\in \hbox{Queries}\qquad\qquad n\in \hbox{Integers}\\
C \in\hbox{Core}  &::= & x\mid n\mid C_1\ \hbox{op}\ C_2\mid []\mid
Q\mid C-\lambda t.f(t)\mid
(C,\lambda t.f(t))\mid \\
&&(\lambda t.f(t),C)\mid  (C_1,C_2)\mid (A,C_1,C_2)\mid (C_1,A,C_2)\mid \\
& &  (C,f,\lambda.f(t))\mid O(C)\mid F(C)\mid M(C)\mid {{C_1}:=
{C_2}}\mid \\ & & C; \mid A(C)\mid R\mid S(C)\mid C_1C_2 \hbox{If
}(C_1)\ C_2\mid \\
& &\hbox{If }(C_1)\ C_2\ C_3\mid \hbox{While }(C_1)\ C_2 \mid C_1\gg
C_2
\end{eqnarray*}
}}\caption{Core of ImpNet in $C$.}\label{f11}
\end{minipage}}\\ \fbox{
\begin{minipage}{12 cm}
{\footnotesize{
\begin{tabular}{ll}
$C-\lambda t.f(t)={\hbox{Lift} (C,\lambda t.f(t))} $
   &$(C,\lambda t.f(t))={\hbox{ApplyLft} (C,\lambda t.f(t))}$  \\
  $(\lambda t.f(t),C)={\hbox{ApplyRit}
(C,\lambda t.f(t))}$ &$(C_1,C_2)=\hbox{Merge}(C_1,C_2)$ \\
$(A,C_1,C_2)=\hbox{MixFst}(A,C_1,C_2)$& $(C_1,A,C_2)=\hbox{MixSnd}(A,C_1,C_2)$ \\
$(C,f,\lambda.f(t))=\hbox{Filter}(C,\lambda.f(t))$&$O(C)= \hbox{Once}(C)$\\
$F(C)=\hbox{MakForwRule}(C)$&$M(C)=\hbox{MakeRule}(C)$\\
$A(C)= \hbox{AddRules}(C)$&$R(C)= \hbox{Register}(C)$ \\
$ S(C)=\hbox{Send}(C)$&$\hbox{If }(C_1)\ C_2=\hbox{If }(C_1)\
\hbox{then }C_2 $\\
$\hbox{If }(C_1)\ C_2\ C_3=  \hbox{If }(C_1)\ \hbox{then }C_2\
\hbox{else }C_3$& $\hbox{While }(C_1)\ C_2 =\hbox{While }(C_1)\
\hbox{do }C_2 $
\end{tabular}
}}\caption{Desugaring of Core Constructs of
Figure~\ref{f11}.}\label{f12}
\end{minipage}}\\ \fbox{
\begin{minipage}{12 cm}
{\footnotesize{
\begin{tabular}{ll}
$C_1\ \hbox{op}\ C_2={(C_1\hookrightarrow \diamond\ \hbox{ op }
C_2)} $ & $n\ \hbox{op}\ C_2={(C_2\hookrightarrow n\ \hbox{
op } \diamond)} $\\
${({C_1}:= {C_2})}={(C_2\hookrightarrow {{C_1}:= {\diamond}})} $ &
$C;={(C\hookrightarrow \diamond;)} $\\
${\hbox{If }(C_1)\ C_2\ C_3 }={(C_1\hookrightarrow \hbox{If
}(\diamond)\ C_2\ C_3 )} $ &
\end{tabular}
}}\caption{Computational Structural Equations of Dynamic
Semantics.}\label{f13}
\end{minipage}}
\caption{Dynamic Operational Semantics for ImpNet.}\label{dyn1}
\end{figure*}

\begin{figure*}
\centering \fbox{
\begin{minipage}{13 cm}
{\footnotesize{
\begin{tabular}{l}
$[]=. $\\
$C_1C_2=C_1\hookrightarrow C_2 $\\
$C_1\hbox{ op }C_2\Longrightarrow  C_1\hbox{ op}_{\hbox{int}} C_2$\\
$\hbox{If }(C_1)\ C_2\ C_3\Longrightarrow C_2$, where $C_1\not = 0$\\
$\hbox{If }(C_1)\ C_2\ C_3\Longrightarrow C_3$, where $C_1 = 0$\\

Left:\\
$<x-\lambda t.f(t)\hookrightarrow C>_{C^s}\ <x\mapsto
(v_1,v_2,\ldots,v_n),\gamma>_{\hbox{VarSts}}\Longrightarrow\
$\\$<(f(v_1),f(v_2),\ldots,f(v_n))\hookrightarrow C>_{C^s}\
<x\mapsto (v_1,v_2,\ldots,v_n),\gamma>_{\hbox{VarSts}}$\\

ApplyLft:\\
$<(x,\lambda t.f(t))\hookrightarrow C>_{C^s}\ <x\mapsto
((v_1,w_1),(v_2,w_2),\ldots,(v_n,w_n)),\gamma>_{\hbox{VarSts}}\Longrightarrow\
$\\$<((f(v_1),w_1),(f(v_2),w_2),\ldots,(f(v_n),w_n))\hookrightarrow
C>_{C^s}$\\ $<x_1\mapsto (v_1,v_2,\ldots,v_n),x_2\mapsto
(w_1,w_2,\ldots,w_n),\gamma>_{\hbox{VarSts}}$\\

ApplyRit:\\
$<(\lambda t.f(t),x)  \hookrightarrow C>_{C^s}\ <x\mapsto
((v_1,w_1),(v_2,w_2),\ldots,(v_n,w_n)),\gamma>_{\hbox{VarSts}}\Longrightarrow\
$\\$<((v_1,f(w_1)),(v_2,f(w_2)),\ldots,(v_n,f(w_n)))\hookrightarrow
C>_{C^s}$\\ $<x_1\mapsto (v_1,v_2,\ldots,v_n),x_2\mapsto
(w_1,w_2,\ldots,w_n),\gamma>_{\hbox{VarSts}}$\\

Merge:\\
$<(x_1,x_2) \hookrightarrow C>_{C^s}\ <x_1\mapsto
(v_1,v_2,\ldots,v_n),x_2\mapsto
(w_1,w_2,\ldots,w_n),\gamma>_{\hbox{VarSts}}\Longrightarrow\
$\\$<((v_1,w_1),(v_2,w_2),\ldots,(v_n,w_n))\hookrightarrow C>_{C^s}$
\\ $<x_1\mapsto (v_1,v_2,\ldots,v_n),x_2\mapsto
(w_1,w_2,\ldots,w_n),\gamma>_{\hbox{VarSts}}$\\

MixFst:\\
$<(A,x_1,x_2) \hookrightarrow C>_{C^s}\ <x_1\mapsto
(v_1,v_2,\ldots,v_n),x_2\mapsto
(w_1,w_2,\ldots,w_n),\gamma>_{\hbox{VarSts}}\Longrightarrow\
$\\$<((A_1,w_1),\dots,(A_n,w_n))=((A\cup\{v_1\},w_1),(A_1\cup\{v_2\},w_2),
\ldots,(A_{n-1}\cup\{v_n\},w_n) )$\\ $\hookrightarrow C>_{C^s}\
<x_1\mapsto (v_1,v_2,\ldots,v_n),x_2\mapsto
(w_1,w_2,\ldots,w_n),\gamma>_{\hbox{VarSts}} $\\

MixSnd:\\
$<(x_1,A,x_2) \hookrightarrow C>_{C^s}\ <x_1\mapsto
(v_1,v_2,\ldots,v_n),x_2\mapsto
(w_1,w_2,\ldots,w_n),\gamma>_{\hbox{VarSts}}\Longrightarrow\
$\\$<((v_1,A_1),\dots,(v_n,A_n))=((v_1,A\cup\{w_1\}),(v_2,A_1\cup\{w_2\}),
\ldots,(v_n,A_{n-1}\cup\{w_n\}) )$\\ $\hookrightarrow C>_{C^s}\
<x_1\mapsto (v_1,v_2,\ldots,v_n),x_2\mapsto
(w_1,w_2,\ldots,w_n),\gamma>_{\hbox{VarSts}} $\\

Filter:\\
$< (C,f,\lambda.f(t)) \hookrightarrow C>_{C^s}\ <x\mapsto
(v_1,v_2,\ldots,v_n),\gamma>_{\hbox{VarSts}}\Longrightarrow\
$\\$<(\ldots,v_i,\ldots\mid f(v_i)=\hbox{ture})\hookrightarrow
C>_{C^s}\
<x\mapsto (v_1,v_2,\ldots,v_n),\gamma>_{\hbox{VarSts}}$\\

Once:\\
$<O(x)\hookrightarrow C>_{C^s}\ <x\mapsto
v,\gamma>_{\hbox{VarSts}}\Longrightarrow\
<(v^1,\ldots,v^n)\hookrightarrow C>_{C^s}\
<x\mapsto v,\gamma>_{\hbox{VarSts}}$ \\

MakForwRule:\\
$< F(C)\hookrightarrow C>_{C^s}\  <x\mapsto
((v^1_1,v^2_1,v^3_1),(v^1_2,v^2_2,v^3_2),\ldots,(v^1_n,v^2_n,v^3_n)),
\gamma>_{\hbox{VarSts}}\Longrightarrow\
$\\$<[(v^1_1,(v^3_1,sendout(v^2_1))),(v^1_2,(v^3_2,sendout(v^2_2))),
\dots,(v^1_n,(v^3_n,sendout(v^2_n)))]$\\$ \hookrightarrow C>_{C^s}\
<x\mapsto ((v^1_1,v^2_1,v^3_1),(v^1_2,v^2_2,v^3_2),
\ldots,(v^1_n,v^2_n,v^3_n)),\gamma>_{\hbox{VarSts}}$ \\

MakeRule:\\
$<M(C) \hookrightarrow C>_{C^s}\ <x\mapsto
((v^1_1,a_1,v^2_1),(v^1_2,a_2,v^2_2),\ldots,(v^1_n,a_n,v^2_n)),
\gamma>_{\hbox{VarSts}}\Longrightarrow\
$\\$<[[(v^1_1,a_1(v^1_2)),(v^2_1,a_2(v^2_2)),\dots,(v^i_n,a_n(v^n_2))]]$\\$
\hookrightarrow C>_{C^s}\ <x\mapsto
((v^1_1,v^2_1,v^3_1),(v^1_2,v^2_2,v^3_2),
\ldots,(v^1_n,v^2_n,v^3_n)),\gamma>_{\hbox{VarSts}}$\\

Assignment:\\
$<{{x}:= {v}}   \hookrightarrow C>_{C^s}\
<\gamma>_{\hbox{VarSts}}\Longrightarrow\ <C>_{C^s}\
<\gamma[x\mapsto v]>_{\hbox{VarSts}}$\\

AddRules:\\
$< A(x) \hookrightarrow C>_{C^s}\ <x\mapsto
[rl,\ldots,rl_n],\gamma>_{\hbox{VarSts}}\ <rl>_{\hbox{RlLst}}\
\Longrightarrow\ $\\$<C>_{C^s}\
<x\mapsto [rl,\ldots,rl_n],\gamma>_{\hbox{VarSts}}<[rl,\ldots,rl_n]
\cup rl>_{\hbox{RlLst}}$\\

Register:\\
$<R \hookrightarrow C>_{C^s}\ <rl>_{\hbox{RlLst}}\
<\sigma>_{\hbox{SwchSts}}\ \Longrightarrow\ <C>_{C^s}\
<[]>_{\hbox{RlLst}}\
<rl\cup \sigma>_{\hbox{SwchSts}}$\\

Send:\\
$<S(C)\hookrightarrow C>_{C^s}\ <x\mapsto
((v^1_1,v^2_1,v^3_1),(v^1_2,v^2_2,v^3_2),\ldots,(v^1_n,v^2_n,v^3_n)),
\gamma>_{\hbox{VarSts}}\ {<h>_{\hbox{History}}} $\\
$ \Longrightarrow\ < C>_{C^s}\ <x\mapsto
((v^1_1,v^2_1,v^3_1),(v^1_2,v^2_2,v^3_2),\ldots,(v^1_n,v^2_n,v^3_n)),
\gamma>_{\hbox{VarSts}}$\\
$ {<v_1^i\mapsto \{(v_2^i,v_3^i),h>_{\hbox{History}}}$\\

$<\hbox{While }(C_1)\ C_2 \hookrightarrow C>_{C^s}\ \Longrightarrow\
<\hbox{If }(C_1)\ [C_2\hbox{While }(C_1)\ C_2\hookrightarrow
C]>_{C^s} $\\
\end{tabular}
}}\caption{Equations and Rules of Dynamic Semantics.}\label{dyn2}
\end{minipage}}
\end{figure*}

Figure~\ref{f11} presents the complete $C$ definitions of
\textit{ImpNet}. Typically $C$ definitions include a single
syntactic category $C$. This category is intended to be a minimal
infrastructure for terms definitions; not to be a parsing or
type-checking tool. Typically in such semantics, sorts  are
equivalent to syntactic categories and operations are equivalent to
productions. Hence algebraic signatures somehow coincide with
context-free notations. In the model $C$, Boolean values are treated
using special integer values.

The concepts of well-structured and well-evaluated computations are
necessary for recognizing start and final configurations of the
dynamic semantics. Using equational reasoning of \textit{ImpNet}'s
dynamic semantics of this section, an \textit{ImpNet} computation
$C$ is \textit{well-structured} if it amounts to a well-structured
event transformers or list of statements in \textit{ImpNet}. A
computation is \textit{well-evaluated} if it amounts to a value
$v\in$ \textit{Values} or to the unit computation ''.''.

Figures~\ref{dyn1} and~\ref{dyn2} present all components of dynamic
semantics of \textit{ImpNet}. Mostly, elements of these figures are
self-describing. Wherever possible, it is preferred to  desugar
constructs of our derived language.

There are some special configurations of the dynamic semantics
presented in this section. For a well-structures computation $C$,
the configuration $<<C>_{C^s},\
    {<.>_{\hbox{SwchSts}}},\ {<.>_{\hbox{VarSts}}},
    \  {<.>_{\hbox{RlLst}}},\\  {<.>_{\hbox{History}}}>$ is the
    \textit{start configuration}. For a "," or a value computation $C$,
the configuration $<<C>_{C^s},\
    {<\sigma>_{\hbox{SwchSts}}},\ {<\gamma>_{\hbox{VarSts}}},
    \  {<rl>_{\hbox{RlLst}}},\\  {<h>_{\hbox{History}}}>$
    is a \textit{final configuration}.

The expression $C_1\hookrightarrow C_2$ denotes processing $C_1$
before processing $C_2$. Therefore $C_2$ is somehow a frozen (from
development) computation until it is its turn. The technique of the
evaluation of the langue \textit{ImpNet} is meant to be captured by
the computation equations.  For example, the conditional statement
schedules calculating the condition first while the branches are
frozen. The most intricate rules of Figure~\ref{dyn2} are that of
event transformers. This is so as most of these rules assume
constrains on muliple event variables.

\begin{figure*}
\centering \fbox{
\begin{minipage}{13cm}
{\footnotesize{
$(\emptyset,\{z\mapsto\{id_1,id_2\},x\mapsto \{((srcport(80),sendall,\_),(inport(1),sendcontroller,\_))\},[])$\\
\textbf{$y=\hbox{MakeRule}(x);$}\\
$(\emptyset,\{z\mapsto\{id_1,id_2\},x\mapsto
\{((srcport(80),sendall,\_),(inport(1),sendcontroller,\_))\},
\\y\mapsto\{(srcport(80),[sendall]),(inport(1),[sendcontroller])\}\},\emptyset)$\\
\textbf{$z=\hbox{Lift} (z,\lambda t.(t,y));$}\\
$(\emptyset,\{z\mapsto\{(id_1,\gamma(y)),(id_2,\gamma(y))\},$\\$x\mapsto
\{((srcport(80),sendall,\_),(inport(1),sendcontroller,\_))\},
\\y\mapsto\{(srcport(80),[sendall]),(inport(1),[sendcontroller])\}\}, \emptyset)$\\
\textbf{$\hbox{AddRules}(z);$}\\
$(\emptyset,\{z\mapsto\{(id_1,\gamma(y)),(id_2,\gamma(y))\},$\\$x\mapsto
\{((srcport(80),sendall,\_),(inport(1),sendcontroller,\_))\},
\\y\mapsto\{(srcport(80),[sendall]),(inport(1),[sendcontroller])\}\},
\{(id_1,\gamma(y)),(id_2,\gamma(y))\})$\\
\textbf{Register;}\\
$(\{(id_1,\gamma(y)),(id_2,\gamma(y))\},\{z\mapsto\{(id_1,\gamma(y)),(id_2,\gamma(y))\},$\\$x\mapsto
\{((srcport(80),sendall,\_),(inport(1),sendcontroller,\_))\},
\\y\mapsto\{(srcport(80),[sendall]),(inport(1),[sendcontroller])\}\},
\emptyset)$\\}} \caption{Program 1.} \label{f4}
\end{minipage}}\\ \fbox{
\begin{minipage}{13cm}
{\footnotesize{ $(\emptyset,\{z\mapsto\{id_1,id_2\}
\},[])$\\
\textbf{$y=\hbox{SourceIps};$}\\
$(\emptyset,\{z\mapsto\{id_1,id_2\},
\\y\mapsto\{(ip_1,pk_1),(ip_2,pk_2)\}\},\emptyset)$\\
\textbf{$y=\hbox{ApplyLft} (y,\lambda t.(t,\hbox{port}(t)));$}\\
$(\emptyset,\{z\mapsto\{id_1,id_2\},$\\
$y\mapsto\{(pr_1,pk_1),(pr_2,pk_2)\}\}, \emptyset)$\\
\textbf{$y=\hbox{Lift} (y,\lambda t.(t,\hbox{switch}(t,z));$}\\
$(\emptyset,\{z\mapsto\{id_1,id_2\},$
\\$y\mapsto\{(id_1,pr_1,pk_1),(id_2,pr_2,pk_2)\}\}, \emptyset)$\\
\textbf{$y=\hbox{MakForwRule}(y);$}\\
$(\emptyset,\{z\mapsto\{id_1,id_2\},
\\y\mapsto\{(id_1,(pk_1,sendout(pr_1)),(id_2,(pk_2,sendout(pr_2)))\}\}, \emptyset)$\\
\textbf{$\hbox{AddRules}(y);$}\\
$(\emptyset,\{z\mapsto\{id_1,id_2\},
\\y\mapsto\{(id_1,(pk_1,sendout(pr_1)),(id_2,(pk_2,sendout(pr_2)))\}\},
\\ \{(id_1,(pk_1,sendout(pr_1)),(id_2,(pk_2,sendout(pr_2)))\})$\\
\textbf{Register;}\\
$(\{(id_1,(pk_1,sendout(pr_1)),(id_2,(pk_2,sendout(pr_2)))\},\{z\mapsto\{id_1,id_2\},
\\y\mapsto\{(id_1,(pk_1,sendout(pr_1)),(id_2,(pk_2,sendout(pr_2)))\}\}, \emptyset)$}}
\caption{Program 2.} \label{f5}
\end{minipage}
}\caption{Static Operational Semantics for Two Control programs in
ImpNet.}
\end{figure*}

\section{Controller Programs}\label{s2}\vspace{-4pt}

This section presents several examples of programs constructed using
the syntax of \textit{ImpNet} (Figure~\ref{f1}). The first example
constructs rules based on information stored in the variable $x$ and
then installs the established rules to flow tables of switches
stored in $z$. This program has the following statements.
\[y = \hbox{MakeRule}(x);\]
\[ z=\hbox{Lift} (z,\lambda t.(t,y));\]
\[\hbox{AddRules}(z);\]
\[Register;\]

The first statement of the program makes a rule for each value of
the event stored in $x$. Then the second statement assigns these
rules to switch IDs in the event stored in $z$. The third statement
stores the rule assignment of $z$ in $ir$ as an initial rule
assignment. The last statement of the program adds the established
rules to the flow tables of switches. Figure~\ref{f4} shows the
static operational semantics of this program using the semantics of
the previous section.

The second example constructs forwarding rules based on source IPs
of arriving packets and then installs the established rules to flow
tables of switch IDs stored in $z$. This program has the following
statements.
\[y=\hbox{SourceIps};\]
\[ y=\hbox{ApplyLft} (y,\lambda t.(t,\hbox{port}(t)));\]
\[y=\hbox{Lift} (y,\lambda t.(t,\hbox{switch}(t,z));\]
\[y=\hbox{MakForwRule}(y);\]
\[\hbox{AddRules}(y);\]
\[Register;\]

The first statement of the program assumes a function
\textit{SourceIps} that returns source IPs of arriving packets and
stores them in the form of an event in $y$. The second statement
transfers event of $y$ into event of pairs of IPs and port numbers
through which packets will be forwarded. The third statement
augments values of event in $y$ with switch IDs from the event
stored in $z$. The fourth statement makes a forward rule for each
value of the event stored in $y$. Then the fifth statement stores
the rule assignment of $y$ in $ir$ as an initial rule assignment.
The last statement of the program adds the established rules to the
flow tables of switches. Figure~\ref{f5} shows the static
operational semantics of this program using the semantics of the
previous section.

Using a sequence of prohibited IPs (stored in the variable $a$), the
third example constructs firewall rules. Firewall rules are
established by first applying \textit{ApplyRit} that applies a map
to get the concerned port numbers. Then the map \textit{Lift} is
used twice in a row to add switch IDs to items in $b$ and to
determine that the action of rules being established is prohibition.
Finally the program installs the established rules to flow tables of
switch IDs stored in $c$. This program has the following statements.

\[a=\hbox{ProhibtedIps};\]
\[b=\hbox{ApplyRit} (a,\lambda t.(t,\hbox{port}(t)));\]
\[b=\hbox{Lift} (b,\lambda t.(t,\hbox{switch}(t,c));\]
\[b=\hbox{Lift} (b,\lambda t.(t,\hbox{prohibt}(t,c));\]
\[b=\hbox{MakRule}(b);\]
\[\hbox{AddRules}(b);\]
\[Register;\]

A fourth example, that switches between actions of forwarding and
dropping depending on source IPs, can constructed as a combination
of the second and third examples above.

The examples shown above demonstrate the importance and robustness
of the programming model developed for software-defined networks in
this paper. For example the third example above implements a very
important concept (Firewalling) of networks in a very compact and
accurate way.

\section{Future Work}
\label{s3a}\vspace{-4pt}

There are many interesting directions for future work. One such
direction is to develop methods for static analysis of network
programming languages. Obviously associating these analyses with
correctness proofs, in the spirit
of~\cite{El-Zawawy11-4,El-Zawawy13-2,El-Zawawy13-1,El-Zawawy11-2},
will have many network applications. Developing denotational
semantics~\cite{El-Zawawy07} for network programs is a very
interesting direction for future research. This will increase the
trust level of these programs. Another direction for future work is
to develop type systems to detect event-errors in the sense of the
work in~\cite{El-Zawawy12-8}. Implementing the problems of indoor
mobile target localization (for wireless sensor
networks)~\cite{Gao13} and that of CSPs search
strategies~\cite{Golemanova13} (for control network programmings)
using \textit{ImpNet} are among interesting directions of future
work.

\section{Conclusion}
\label{s4}\vspace{-4pt}

Software-Defined Network (SDN) is a recent architecture of networks
in which a controller device is used to program other network
devices (specially switches). This is done via a sequence of
installing and uninstalling of rules to memories of these devices.

In this paper, we presented a high-level imperative network
programming language, called \textit{ImpNet}, to facilitate the job
of controller. \textit{ImpNet} produces efficient, yet simple, and
powerful programs. \textit{ImpNet} has the advantages of simplicity,
expressivity, propositionally, and more importantly being
imperative. The paper also introduced two concrete operational
semantics to meanings of \textit{ImpNet} constructs. One of the
proposed semantics is static and the other is dynamic. Detailed
examples of using \textit{ImpNet} and the operational semantics were
also illustrated in this paper.

The proposed language can be used to program many network
applications like switch load-balancing. The proposed language  can
also be realized as a new framework for network-programming that
enables applying static and dynamic analysis techniques to network
programs.

\end{document}